# Tetrahedra and Relative Directions in Space Using 2 and 3-Space Simplexes for 3-Space Localization

Kevin T. Acres and Jan C. Barca

*Monash Swarm Robotics Laboratory*
*Faculty of Information Technology*
*Monash University, Victoria, Australia*

# Abstract

This research presents a novel method of determining relative bearing and elevation measurements, to a remote signal, that is suitable for implementation on small embedded systems – potentially in a GPS denied environment. This is an important, currently open, problem in a number of areas, particularly in the field of swarm robotics, where rapid updates of positional information are of great importance. We achieve our solution by means of a tetrahedral phased array of receivers at which we measure the phase difference, or time difference of arrival, of the signal. We then perform an elegant and novel, albeit simple, series of direct calculations, on this information, in order to derive the relative bearing and elevation to the signal. This solution opens up a number of applications where rapidly updated and accurate directional awareness in 3-space is of importance and where the available processing power is limited by energy or CPU constraints.

# 1. Introduction

Motivated, in part, by the currently open, and important, problem of GPS free 3D localisation, or pose recognition, in swarm robotics as mentioned in (Cognetti, M. et al., 2012; Spears, W. M. et al. 2007; Navarro-serment, L. E. et al.1999; Pugh, J. et al. 2009), we derive a method that provides relative elevation and bearing information to a remote signal. An efficient solution to this problem opens up a number of significant applications, including such implementations as space based X/Gamma ray source identification, airfield based aircraft location, submerged black box location, formation control in aerial swarm robotics, aircraft based anti-collision aids and spherical sonar systems.

A certain amount of prior work in this area does exist and is well documented, particularly in (Bishop, A. et al. 2008) where the authors present a novel derivation of the traditional maximum likelihood estimation (TML) method of localisation. This paper further categorises a number of other localisation methods, such as *Traditional Maximum Likelihood*, *Closed-Form Linear Least Squares* and *Constrained Weighted Linear Least Squares*. Other works present direct solutions to GPS, or hyperbolic, type localisation equations such as those by (Krause, L. 1987; Chan, Y. T. & Ho, K. C. 1994).

A common theme through various localisation schemes, such as those documented above, is in the amount of processing required. For example convergence of least squares algorithms, matrix arithmetic and the solving of a series of simultaneous equations are all reasonably complex and potentially time consuming procedures to perform by software. Conversely, the system presented in this document shows that elevation and bearing information can be derived, in a fast and simple manner, by making use of the mathematical properties of regular 2-space and 3-space simplexes (or equilateral triangles and tetrahedrons). This permits us to efficiently derive localisation information by a series of direct calculations involving just trigonometric and square root functions.

The design presented here is facilitated by providing an array of transducers, positioned at the vertices of a regular tetrahedron, which supply time difference of arrival information to a four channel receiver. This design is purposely agnostic of the type of transducer used. In the absence of ideal isotropic transducers it should be noted that, for most applications, we are mainly interested in the time (or phase) difference of arrival information of a signal and not its amplitude component.

The remainder of this paper is outlined as follows. In section 2 we present a high level overview of the system. Section 3 presents some high impact applications for this work. In Section 4 we look at the mathematical basis of this work. Section 5 then examines the near and far field effects as they apply to this work. Section 6 looks at the symmetrical properties of tetrahedrons with a view to error minimisation. Section 7 examines degraded modes of operation. In Section 8 we present mathematical simulations over a range of distances. Finally, we conclude the document in Section 9.

## 2. System Overview

This document presents a method for deriving localisation information from time difference of arrival (TDOA) information received at an array of antennas, or transducers, that are positioned such that they form the vertices of a regular tetrahedron. The choice of tetrahedron is made since the location of a point in 3 space requires that 4 measurements are made. This follows from the general condition that any point in $n$-space requires $n+1$ measurement points in order to uniquely identify its location.

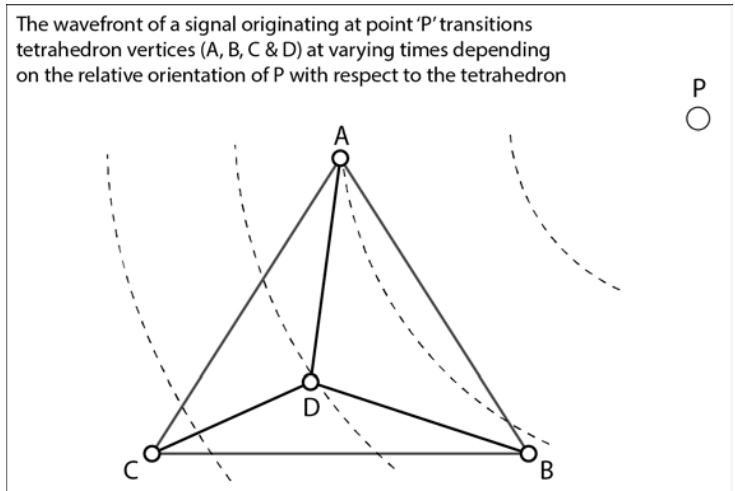

Figure 1  Signal Time Difference of Arrival at Tetrahedron Vertices

Signals received, by the array, are communicated, individually through a suitable medium, where the time delay for each signal in transit is equal, to an array of 4 receivers at which the TDOA information may be extracted. Alternately, time division multiplex (TDM) techniques may be used with a single receiver. In this case the four signals are multiplexed at the receiver input and de-multiplexed at the phase (or TDOA) comparison stage.

Then, by utilising the proven mathematical properties of 2 and 3 space simplexes (equilateral triangle and regular tetrahedron), we are able to derive both elevation and bearing information, relating to the received signal, in a short series of directly calculated steps. We are able to further utilise the properties of a 2 space simplex in order to provide backup elevation information in the event of degradation of one of the four signals (antenna shielding for example).

The novelty of this system is that the mathematical processing is minimal, allowing faster update times, for a given amount of processing power, than would be possible with previously published algorithms.

# 3. High Impact Applications

A broad range of high impact applications exist for this direction finding technology. A selected subset of these applications is presented below.

### 3.1. Space Based X/Gamma Ray Source Identification

As an alternate to space based 'Pair Telescopes' (Michelson, 2003) a tetrahedral array of gamma ray detectors could be utilised in order to derive the direction of a gamma ray burst. Similarly with X-rays.

A large tetrahedral array would be constructed with, for example, either gas filled X-ray detectors or spark chamber Gamma ray detectors at the vertices. Such a system is expected to provide a cost effective adjunct to currently deployed systems. Its utility derives from the system's ability to ascertain the direction of a burst, rather than that of individual high energy particles.

### 3.2. Airfield Based Aircraft Location

To help with pilot guidance into an airport, with possibly no other form of navigation assistance but with a manned control tower, a small array can be equipped with 4 phased VHF receivers. The size of the array would be such that it would take advantage of the ~2.5 metre wavelength used for general aviation radio communication; permitting an array with antenna separation of about one metre which would be sufficient for good direction resolution. The direction finding capabilities can be further enhanced by altitude information, given verbally by the pilot, in order to determine both direction and distance of his aircraft from the airport.

### 3.3. Locating Submerged Black Boxes

A phased array of 'ping' detectors would be an ideal solution for submerged black box location. It would no longer be necessary to search for the strongest signal; rather each received ping would arrive from a known direction, allowing the rapid convergence of any search. As sound travels relatively slowly (as compared to radio signals) the array can be fairly compact and still deliver good directional resolution.

### 3.4. Aerial Swarm Robotics

The 'direct calculation' of this method facilitates use by low end and energy constrained embedded systems, such as used by drones that may form part of an aerial swarm. In this instance we can make use of, say, a TV transmitter on a lead drone and detect the phase difference, and hence the time difference of arrival (TDOA), of an unused sub-carrier such as would be used for one of the two sound channels. This information can be used by other members of the swarm to help maintain their positioning in the swarm. In this application it is important to also have accurate distance measurements. This can be achieved by a number of means. For example, exchanging altitude information permits determination of distance by means of trigonometry. Potentially, low power 'Ultra Wide Band' ranging chipsets may also be used to directly derive the distance.

The following two diagrams illustrate the use of two sound channels associated with a modular TV transmitter of the type commonly used for first person view (FPV) video information often used in radio controlled aircraft. In this example, one sound channel carrier is used solely for vector determination, whereas the other is used as a telemetry channel for communicating altitude information.

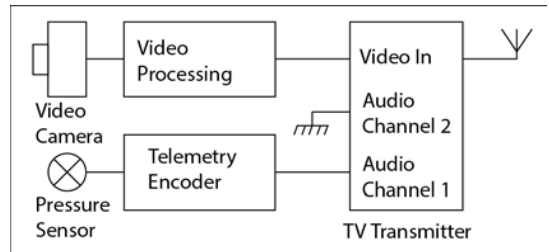
**Figure 2  Localisation Transmitter**

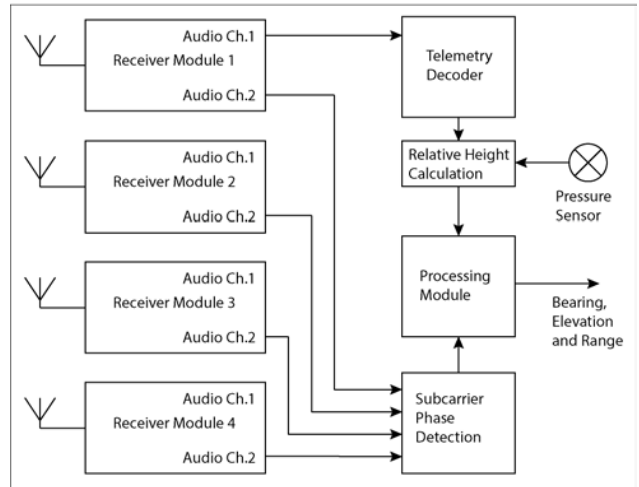
**Figure 3  Localisation Receiver**

Ranging information may then be calculated from the two known values of the height difference and the angle of elevation from one drone to the other.

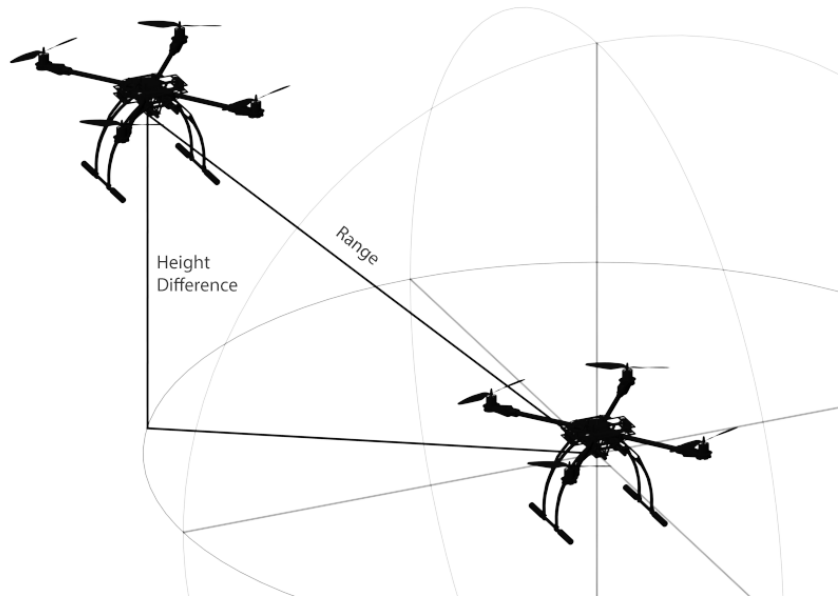

**Figure 4 Deriving Range between two Drones**

In this instance, let $\theta$ = angle of elevation, $h$ = height difference then the range $R$ may be calculated as $R = h/\sin(\theta)$.

### 3.5. Aircraft Based Anti-Collision Aid

A reduced complexity configuration is acceptable in the instance where ambiguity exists in one plane. In this case a triangular antenna array may be used and the ambiguity resolved via another means. Take for example the case of an aircraft fitted with a triangular array, with antennas mounted at the tail and at each wing tip. The system can now be used to unambiguously derive bearing. However, the elevation becomes ambiguous without use of a 4$^{th}$ antenna/receiver.

Fortunately, this ambiguity may easily be resolved by means of a transfer of altimeter data. If need be this can be carried as telemetry on the transmitted signal. The difference in the received altimeter data when compared with that of the receiving aircraft would be sufficient to resolve the ambiguity in elevation.

### 3.6. Spherical Sonar System

A novel application for a tetrahedral 'Time Difference of Arrival' (TDOA) array is in a spherical sonar system. In this instance the receiving transducer spacing is designed such that it is greater than the pulse width of the sonar transmitter. The sonar transmitting transducer is preferably located at the centroid of the tetrahedral array and sends out periodic pulses; the power, width and spacing of which are the major factors in the range and resolution of the system.

The receiving system, rather than determine phase differences in the incoming signal, sends incoming data to a series of four buffers in memory. These buffers are then examined for correlation of both amplitude and time of arrival of incoming pulse reflections. From this, a 3D map can be created of objects that return a signal from the transmitted pulse.

The following diagram outlines the major components of such a system:

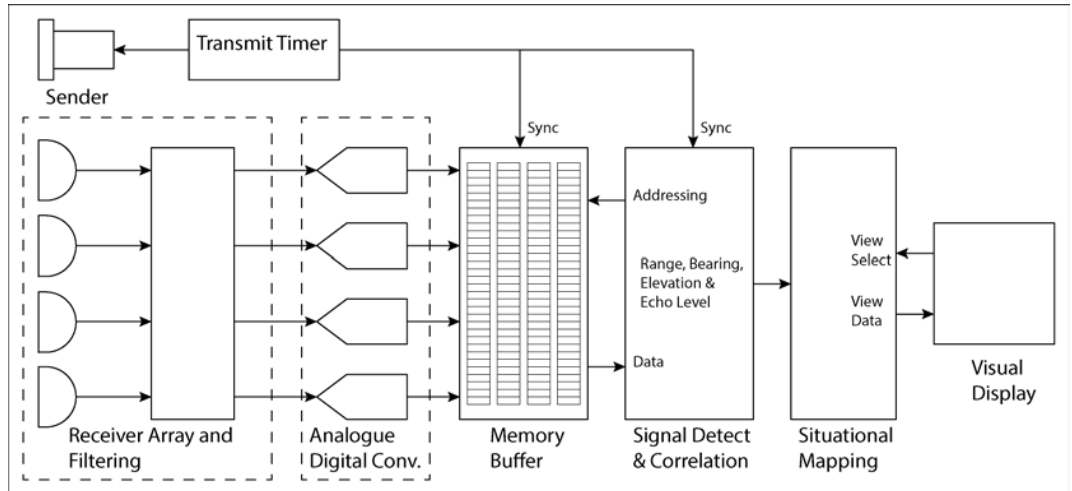

Figure 5  Outline of Spherical 3D Sonar System

## 4. Mathematical Basis

The majority of this work is either based upon or derived from the following theorem by Apostol and Mnatsakanian. Whilst it specifically relates to points in 2-space the results do, however, also hold for 3-space and higher dimensions (Apostol & Mnatsakanian, 2001). In our instance, for operations in 3-space, it is sufficient to substitute a sphere for a circle.

Theorem 1: *Given n fixed points in a plane, a point moving in the plane of these points in such a way that the sum of the squares of the distances from the points is constant traces out a circle whose center is at the centroid of the fixed points* (Apostol & Mnatsakanian, 2001).

### 4.1. Mathematical Modelling

We now take a look at some modelling in which we first present useful equalities for switching between spherical and edge measurements. This is followed by a step by step derivation of the equations, for both regular tetrahedra and equilateral triangles, on which this work is based .

#### 4.1.1. Useful Equalities

To facilitate operations on both tetrahedra and triangles use will be made of the length of an edge rather than the radius of the circumscribed sphere or circle.

In the case of a regular tetrahedron $R$ denotes the radius of the circumscribed sphere, $l$ the length of an edge and $t$ represents the height of the tetrahedron; then

$$t = \frac{\sqrt{6}}{3}l$$

$$R = \frac{3}{4}t = \frac{3}{4} * \frac{\sqrt{6}}{3}l = \frac{\sqrt{6}}{4}l$$

In particular we are interested in representing $R^2$ in terms of $l$

$$R^2 = (\frac{\sqrt{6}}{4}l)^2 = \frac{3}{8}l^2$$

It follows that the sum of squares of distances from the centroid to the vertices of a regular tetrahedron will then have the following equality

$$4R^2 \equiv \frac{4*3}{8}l^2 \equiv 1.5l^2 \tag{1}$$

In the case of an equilateral triangle we consider the case where $R$ is the radius of the circumscribed circle and $l$ the length of an edge

$$R = \frac{l}{\sqrt{3}}l$$

For an equilateral triangle we may show an equality of the sum of squares from the centroid to the vertices as

$$3R^2 \equiv 3 * \left(\frac{l}{\sqrt{3}}l\right)^2 \equiv 3 * \frac{l^2}{3} \equiv l^2 \tag{2}$$

.
### 4.1.2. Regular Tetrahedron

The regular tetrahedron is a regular 3-space simplex with 6 equal sized edges and 4 vertices. Importantly it has 12 rotational symmetries of which use will later be made of 4. Essential to this presentation are the properties of the sum of squares of distances from any given point in 3-space to each of the vertices of a regular tetrahedron. In order to demonstrate the required properties we derive them algebraically below. Initially we will determine each of the vertices $a, b, c$ and $d$ of a regular tetrahedron as a function of radius $R$ of the circumscribing sphere as shown in Table 1 below.

**Table 1 Vertex Placement as a Function of the Radius of a Circumscribing Sphere**

| Vertex | X coordinate | Y coordinate | Z coordinate |
|---|---|---|---|
| a | 0 | $2\sqrt{2R^2}/3$ | $-R/3$ |
| b | $\sqrt{6}R/3$ | $-\sqrt{2R^2}/3$ | $-R/3$ |
| c | $-\sqrt{6}R/3$ | $-\sqrt{2R^2}/3$ | $-R/3$ |
| d | 0 | 0 | $R$ |

We now place an arbitrary remote point

$$P = [px, py, pz]$$

where distance $x$ from $P$ to the centroid of the tetrahedron is

$$x = \sqrt{px^2 + py^2 + pz^2}$$

To help simplifying our construction we perform an algebraic substitution for the square roots. Let $S = \sqrt{6}$ and $T = \sqrt{2R^2}$, then we calculate $r_a = P - a$, $r_b = P - b$, $r_c = P - c$, $r_d = P - d$

where

$$r_a = P - a = [px, py, pz] - [0, \tfrac{2}{3}T, -\tfrac{1}{3}R] = [px, py - \tfrac{2}{3}T, pz + \tfrac{1}{3}R]$$
$$r_b = P - b = [px, py, pz] - [\tfrac{1}{3}S.R, -\tfrac{1}{3}T, -\tfrac{1}{3}R] = [px - \tfrac{1}{3}S.R, px + \tfrac{1}{3}T, px + \tfrac{1}{3}R]$$
$$r_c = P - c = [px, py, pz] - [-\tfrac{1}{3}S.R, -\tfrac{1}{3}T, -\tfrac{1}{3}R] = [px + \tfrac{1}{3}S.R, px + \tfrac{1}{3}T, px + \tfrac{1}{3}R]$$
$$r_d = P - d = [px, py, pz] - [0, 0, R] = [px, py, pz - R]$$

We proceed by deriving the individual squares of distances

$$r_a^2 = px^2 + py^2 - \tfrac{4}{3}T.py + pz^2 + \tfrac{2}{3}R.pz + \tfrac{4}{9}T^2 + \tfrac{1}{9}R^2$$
$$r_b^2 = px^2 - \tfrac{2}{3}S.R.px + py^2 + \tfrac{2}{3}T.py + pz^2 + \tfrac{2}{3}R.pz + \tfrac{1}{9}T^2 + \left(\tfrac{1}{9}S^2 + \tfrac{1}{9}\right)R^2$$
$$r_c^2 = px^2 + \tfrac{2}{3}S.R.px + py^2 + \tfrac{2}{3}T.py + pz^2 + \tfrac{2}{3}R.pz + \tfrac{1}{9}T^2 + \left(\tfrac{1}{9}S^2 + \tfrac{1}{9}\right)R^2$$
$$r_d^2 = px^2 + py^2 + pz^2 - 2R.pz + R^2$$

The sum of squares of distances is then

$$r_a{}^2 + r_b{}^2 + r_c{}^2 + r_d{}^2$$

which expands to

$$4px^2 + 4py^2 + 4pz^2 + \tfrac{2}{3}T^2 + \left(\tfrac{2}{9}S^2 + \tfrac{4}{3}\right)R^2$$

Substituting for $S^2 = 6$ and $T^2 = 2R^2$

$$4px^2 + 4py^2 + 4pz^2 + \tfrac{2}{3}2R^2 + \left(\tfrac{12}{9} + \tfrac{4}{3}\right)R^2 = 4(px^2 + py^2 + pz^2) + 4R^2$$

With $x$ as the distance from the tetrahedron centroid to the remote point and $s$ as the sum of squares of distances from the remote point to the tetrahedron vertices

$$x^2 = px^2 + py^2 + pz^2$$

$$s = r_a{}^2 + r_b{}^2 + r_c{}^2 + r_d{}^2 = 4x^2 + 4R^2 \tag{3}$$

For our application, we don't have access to distance $x$ between the centroid and point $P$. So initially we recreate our sum of squares Equation (3) whilst removing $x$ from it via

$$(r_a - x)^2 + (r_b - x)^2 + (r_c - x)^2 + (r_d - x)^2$$

which expands to

$$4x^2 - 2(r_a + r_b + r_c + r_d)x + (r_a{}^2 + r_b{}^2 + r_c{}^2 + r_d{}^2) \tag{4}$$

However, we know from Equation (3) that

$$(r_a{}^2 + r_b{}^2 + r_c{}^2 + r_d{}^2) \equiv 4x^2 + 4R^2 \tag{5}$$

By assigning

$$r_1 = r_a - x, \ r_2 = r_b - x, \ r_3 = r_c - x, \ r_4 = r_d - x$$

then substituting for Equation (5) and $r_{1-4}$

$$8x^2 - 2(r_1 + r_2 + r_3 + r_4 + 4x)x + 4R^2 = r_1^2 + r_2^2 + r_3^2 + r_4^2 \tag{6}$$

which expands to

$$2(r_1 + r_2 + r_3 + r_4 + 4x)x$$

We get

$$8x^2 + 2(r_1 + r_2 + r_3 + r_4)x$$

This simplifies Equation (6) to

$$-2(r_1 + r_2 + r_3 + r_4)x + 4R^2 = r_1^2 + r_2^2 + r_3^2 + r_4^2$$

which is equivalent to

$$-2(r_1 + r_2 + r_3 + r_4)x = r_1^2 + r_2^2 + r_3^2 + r_4^2 - 4R^2$$

By multiplying both sides by $-1$

$$2(r_1 + r_2 + r_3 + r_4)x = 4R^2 - (r_1^2 + r_2^2 + r_3^2 + r_4^2)$$

and solving for $x$

$$x = \frac{4R^2 - (r_1^2 + r_2^2 + r_3^2 + r_4^2)}{2(r_1 + r_2 + r_3 + r_4)} \qquad (7)$$

We have demonstrated that $x$ can be recovered solely from the measurements from which it has been removed. Equation (7) therefore becomes central to our discussion. Having discarded the original distance of the remote point $P$ and as $P$ approaches infinity the $2(r_1 + r_2 + r_3 + r_4)$ component approaches $0$. We utilize this phenomenon in a 'time difference of arrival' application by applying an equal offset to each $r_k$ ensuring that $\sum_{k=1}^{4} r_k = 0$ to effectively place $P$ at infinity, whilst retaining its angular properties with respect to the centroid. (Note that time difference of arrival measurements imply that at least one of $r_1, r_2, r_3, r_4 = 0$. Failing this, the minimum $r_k$ first needs subtracting from all $r_k$). This is performed by the following procedure in which $o$ is set to the required offset

$$o = \frac{(\sum_{k=1}^{4} r_k)}{4}$$

where

$$r_1 = r_1 - o$$
$$r_2 = r_2 - o$$
$$r_3 = r_3 - o$$
$$r_4 = r_4 - o$$

This transform has an extra side effect insomuch as the following now holds; assuming that $R$ is the radius of the circumscribed sphere of the tetrahedron and $b$ an angle related to the remote point, then

$$b = \arccos\left(\frac{r_x}{R}\right) \qquad (8)$$

Importantly, for any vertex the angle to the remote point $P$ is $90 - \arccos\left(\frac{r_x}{R}\right)$ degrees relative to the plane normal to a line from the vertex to the centroid. Positive values are above the plane and negative values below (with respect to the vertex). Except for near field effects, this is a direct reading relative the related plane and is not affected by changes in bearing. A corollary to this is that, in the case of a regular tetrahedron sitting with one face on the horizontal plane, that the $r_x$ value at the upper vertex may be used directly to derive the elevation of a remote point.

Note that the closer the distance from the centroid to $p$ the more a 'near field' effect on the vertices of the tetrahedron becomes apparent. Conversely, as $P$ moves further away this effect decreases. For near field measurements, checks and supplementary offsets are required to avoid sine and cosine limits.

### 4.1.3. Equilateral Triangle

The equilateral triangle is a regular 2-space simplex with 3 equal sized edges and 3 vertices.

Theorem 1 also applies here and in this case, the sum of squares $s$, assuming that $l$ is the length of a side, $x$ the distance from the centroid to the point $p$ and $R$ the radius of the circumscribed circle, may be represented thus; bearing in mind the equality at (2)

$$s = 3x^2 + 3R^2 \equiv 3x^2 + l^2$$

In a similar fashion to the tetrahedron we can derive distance $x$ to a remote point via

$$x = \frac{3R^2 - (r_1^2 + r_2^2 + r_3^2)}{2(r_1 + r_2 + r_3)} \equiv \frac{l^2 - (r_1^2 + r_2^2 + r_3^2)}{2(r_1 + r_2 + r_3)} \qquad (9)$$

In contrast to the tetrahedron vertices, which are used to measure elevation, we use one, or possibly more, of the triangle faces to measure the bearing to the remote point. Not only does this require that

$$r_1 + r_2 + r_3 = 0 \qquad (10)$$

but also that

$$(r_1^2 + r_2^2 + r_3^2) = 1.5R^2 \qquad (11)$$

In particular, conformance to Equation (11) ensures that near field distortion is minimized with the resulting reduction in the measurement error component.

To prepare a triangle for measurement of bearing information we follow these steps:

1. Subtract the smallest of $r_1, r_2, r_3$ from each of $r_1, r_2, r_3$
2. Subtract the average of $r_1, r_2, r_3$ from each of $r_1, r_2, r_3$
3. Scale values $r_1, r_2, r_3$ such that $(r_1^2 + r_2^2 + r_3^2) = 1.5R^2$
4. Read the bearings at each $r_k$, $\alpha = \frac{\cos(r_k)}{R}$

The scaling function in Step 3 ensures that the triangle face becomes effectively in the same plane as the remote point. In this scaling algorithm[1] point $P = (r_1, r_2, r_3)$ and unit normal $N = (1,1,1)/\sqrt{3}$, while a projected point Q is derived using

$$Q = P - (P.N)N \qquad (12)$$

We then scale $Q$ to derive point $O$ via

$$O = \frac{\sqrt{1.5R^2}Q}{\|Q\|} \qquad (13)$$

---

[1] This algorithm originates from a discussion at http://math.stackexchange.com/questions/1200676/how-to-resolve-abc-0-and-a2b2c2-d and was authored by http://math.stackexchange.com/users/31744/bubba .

Point $O$ will now conform to the constraints of Equation (10) and (11), and the resulting values can be used directly for bearing derivation.

Assuming $S$ is our scaling factor and $\theta$ the angle of incidence of a line between the centre of the triangle and the remote point, the angle of the remote point relative to the plane of the triangle can be obtained as follows

$$S = \frac{\sqrt{1.5R^2}}{\|Q\|}, \theta = \arccos\left(\frac{1.5}{S}\right) \tag{14}$$

This information can be used as a check for the elevation derived from the tetrahedron as shown at Equation (8), assuming that near field effects do not cause $S$ to drop below 1.5, in which case an elevation of 0 degrees must be assumed. This information is also useful in degraded, or intentional 'reduced complexity' modes of operation, albeit with ambiguity as to whether the remote point is above or below the plane of the triangle. This ambiguity may, however, be easily resolved by 'out of band' methods, for example by comparing (or inferring) altitude information.

## 5. Discussion on Near and Far Fields

In the context of this document we diverge a little from the usual meaning of near and far fields. Here we refer to 'Far Field' as being a signal source far enough away such that its wave front is essentially flat as it passes through the measurement points at the vertices of the tetrahedron. Conversely, 'Near Field' signal sources will have a noticeable and measurable curvature to the wave front, which in this design can lead to a certain, but predictable, amount of error in the measurements. The near field effect can cause the time difference of arrival of the signals to be either closer or further apart than expected depending solely on the relative orientation of the signal and tetrahedral array.

In the following sections we examine the near field effect in 2 dimensions, as it would apply, for example, to the triangle that forms the base of the tetrahedron. We examine scenarios where the remote signal is at 0, 30 and 60 degrees orientation relative to, and on the same plane as, this triangle.

### 5.1. Signal at 0 Degrees Relative to Apex and Centroid of Triangle

With the signal source oriented along a line through the centroid and apex of the triangle we have a situation where the signal travels a longer distance to the further receivers than would be expected if the source was at a point at infinity. This effectively increases the time difference of arrival between the receiver at point $A$, in the diagram below, and points $B$ and $C$. In this instance this effect is symmetric at $B$ and $C$ with the result that the derived bearing is not affected.

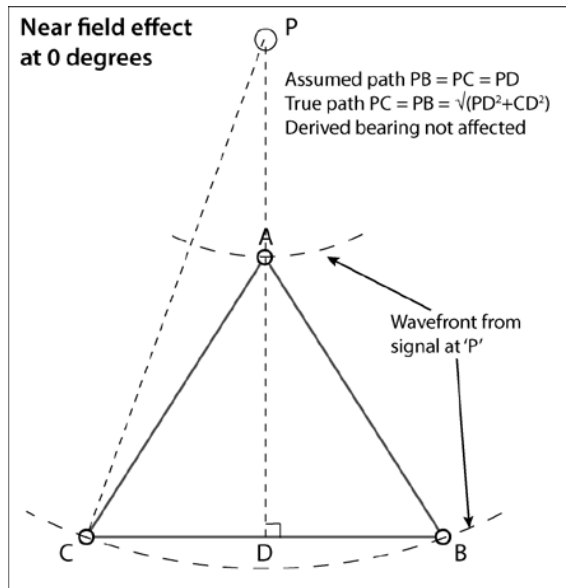
Figure 6  Near Field Effect at 0 Degrees

### 5.2. Signal at 30 degrees relative to apex and centroid of triangle

As shown in the diagram below, the situation is a little more complex for the 30 degree example. In this instance the perceived signal delay at $B$ has the effect of making the signal source at $P$ seem a little further away relative to the delays at $A$ and $C$. This displaces the apparent angular position of $P$ to less than the expected 30 degrees.

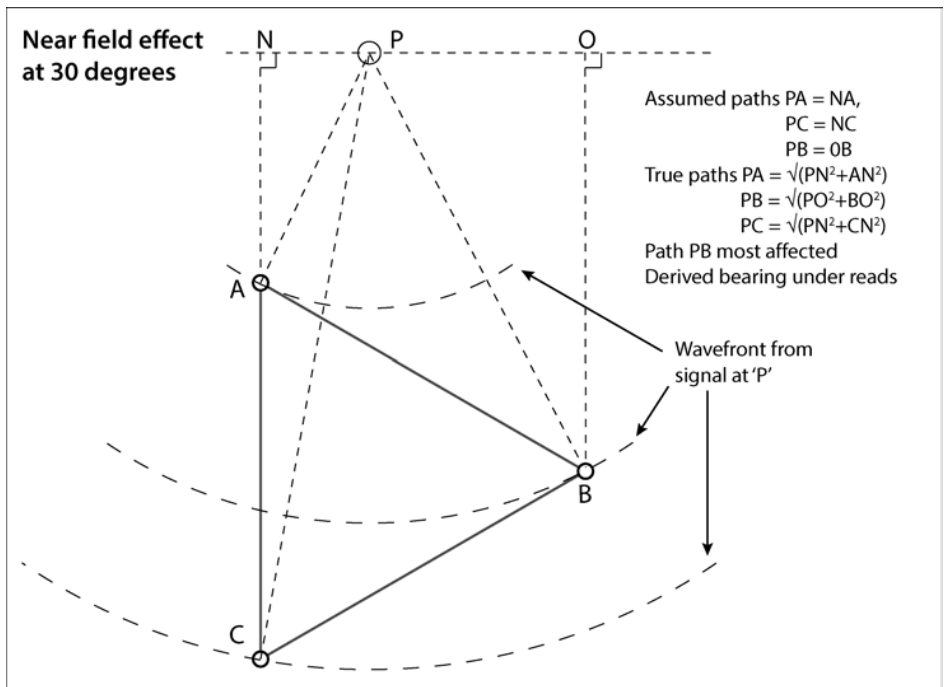
Figure 7  Near Field Effect at 30 Degrees

### 5.3. Signal at 60 degrees relative to apex and centroid of triangle

In this example, pictured below, with a bearing of 60 degrees, the signal wave front at *A* and *B* is delayed relative to that which would be expected for a signal from a point at infinity. Overall the time difference of arrival of the signal at points *A* and *B* compared with *C* is less than would have otherwise have been expected. However, in common with the 0 degree example, because the effect is symmetric it does not affect the derived bearing.

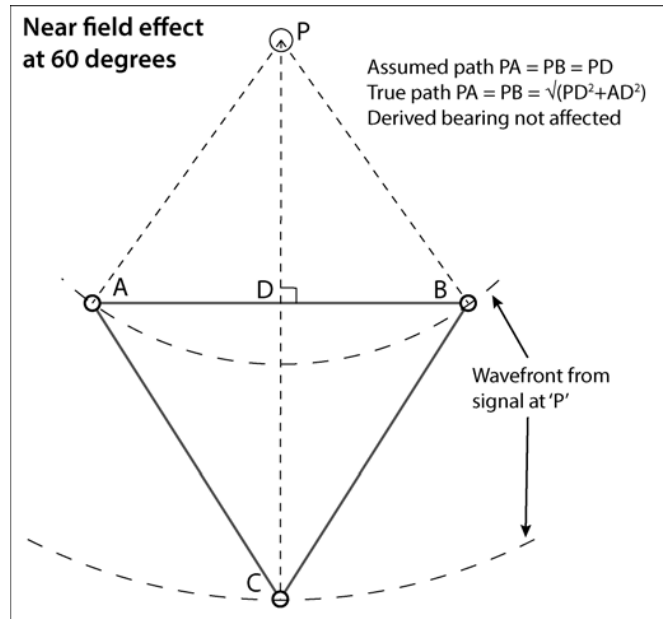

Figure 8  Near Field Effect at 60 Degrees

## 6. Use of Tetrahedral Symmetries for 'Best Range' Measurements

In order to help maintain the best possible resolution, in elevation measurements, these are limited to $\pm 35.5°$; this being just larger than one half of the tetrahedron dihedral angle of $\arccos(1/3)$ or $\approx 70.53°$. Irrespective of the orientation of the tetrahedron there will always be at least one vertex/centroid line that has an angle within the range of $\pm \frac{\arccos(1/3)}{2}$ degrees from the signal source relative to its normal[2]. This knowledge permits us to select a symmetrical orientation of the tetrahedron relative to the signal source in which the elevation is within the range $\pm 35.5°$. This allows operations on shallow angles when determining the bearing component of the signal and is of great benefit when operations that rely on sines and/or cosines are carried out.

The following image shows the coverage of a sphere by four $\pm 35.5°$ spherical segments applicable to each vertex of a regular tetrahedron.

---

[2] This has a proof in a discussion at http://math.stackexchange.com/questions/1279388/ which was authored by http://math.stackexchange.com/users/229191/san .

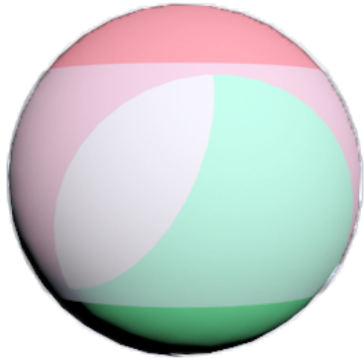

**Figure 9  Coverage of a sphere by four +/- 35.5 degree segments**

## 7. Degraded Modes of Operation

In this section we take a look at contingencies for degraded modes of operation.  In particular we look at a method of handling multipath reflections and antenna shielding.

### 7.1. Multipath Reflection Mitigation

Multipath reflections have the potential to cause inaccurate readings.  There a number of ways that this can be addressed.  This system works entirely on phase difference or, equivalently, time difference of arrival, of a signal, at the 4 receivers, which lends itself to the possibility of a certain amount of signal conditioning.  In particular limiting at an intermediate frequency (IF) stage allows use of the 'FM capture effect' (Leentvaar & Flint, 1976) which reduces the effect of an interfering signal.

A second method of addressing this problem is by gating the incoming signals. This can be especially useful when the antenna separation is small when compared to the wavelength of the signal being received.  This works by waiting for the first event, be it zero crossing or pulse.  Input from the remaining receivers is then enabled for a time equal to the longest transition time of a signal across the array.  Following this time delay the receivers are all disabled until just before the next event is expected, when they are re-enabled once more.

### 7.2. Antenna Shielding

It is possible that, at some orientations, one of the antennas of the array will become shielded from the remote signal. This may be, for example, as a consequence of the construction of the device around which the antenna array is built.  In this instance we are able to take advantage of Equation (14) and derive elevation information directly from the readings at the remaining three antennas; bearing in mind that this will lose resolution at higher angles and that the special case of 90 degrees needs to be handled separately.

## 8. Simulations and Results

In this section we look at some simulations that allow us to derive the error component of elevation and bearing readings with respect to the distance to the signal. This is important as it provides a baseline for the theoretical limits of the system. However, it is worth noting that practical limits are imposed by measurement resolution and construction tolerances and are expected to be less than the theoretical maximum.

For the purposes of this simulation we model a regular tetrahedron with edge length of 0.5 metres. These dimensions were chosen as they would be suitable for implementation in an aerial swarm robotics context, where the size of the quad rotor drones would permit a relatively unobtrusive antenna array to be built around the main chassis.

Measurements of the error components are made at 1, 10 and 100, 1,000, 100,000 and 1,000,000 meters. This provides a sufficient selection of distances from which the behavioural properties of this system may be determined.

For the purposes of this simulation we restrict the angle of elevation from -35 to +35 degrees, which is approximately half the dihedral angle of a regular tetrahedron of $\approx 70.5°$ (Jackson & Weisstein, 2015) Doing so provides us with better resolution than would have been obtainable at higher angles and takes advantage of the symmetries of a tetrahedron that ensures that any given remote point will always have an angle of incidence, $\theta$, such that $|\theta| <= \arccos(1/3)/2$, or about 35.26 degrees, to at least one plane centred at the centroid and normal to a vertex/centroid line.

Positional errors are calculated individually for the bearing and elevation measurements. This is achieved by assuming that the true and derived bearings form two points on an isosceles triangle, $C$ and $D$ respectively, and that the measurement point, $A$, makes the third, with $\propto = \angle CAD$. The line $CD$ is then bisected by a line $AB$ to form two right angle triangles.

The, two dimensional, error component then becomes:

$$CD = 2BD = 2AD \sin \propto/2 \tag{15}$$

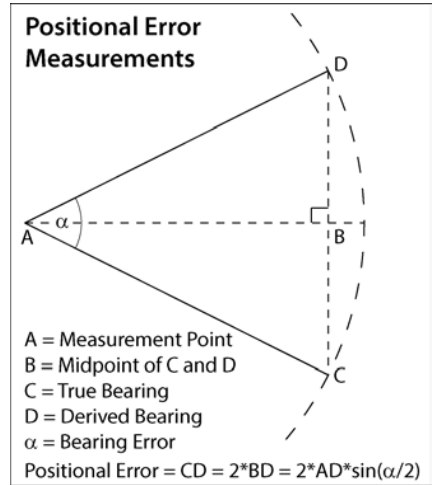

**Figure 10 Positional Error Derivation**

To calculate the positional error in 3-space, the individual error components are derived, $CD$ for the bearing measurement and $C'D'$ for the elevation measurement. The total, absolute, error is then derived by Pythagoras' theorem:

$$Error\ Component = \sqrt{CD^2 + C'D'^2} \qquad (16)$$

Simulation results are plotted below on a series of 3D graphs. In these the $x$ axis represents degrees of bearing, from 0 to 360. The $z$ axis represents elevation from $-35$ through to $+35$ degrees and the $y$ axis represents a suitably scaled error component.

In all simulations below, measurements have been made of the bearing and elevation errors whilst stepping through 0 to 360 degrees of bearing and at elevations from -35 to +35 degrees, both of which have been made in 5 degree increments.

### 8.1. Results at 1 Metre

A simulation was set up for an effective target distance of 1 metre from the centroid of the tetrahedron. Simulated measurements were then made and results saved as a series of excel files from which graphs were plotted for both the bearing and elevation error components. Note that both elevation and bearing errors are given in degrees whereas positional errors are given in centimetres.

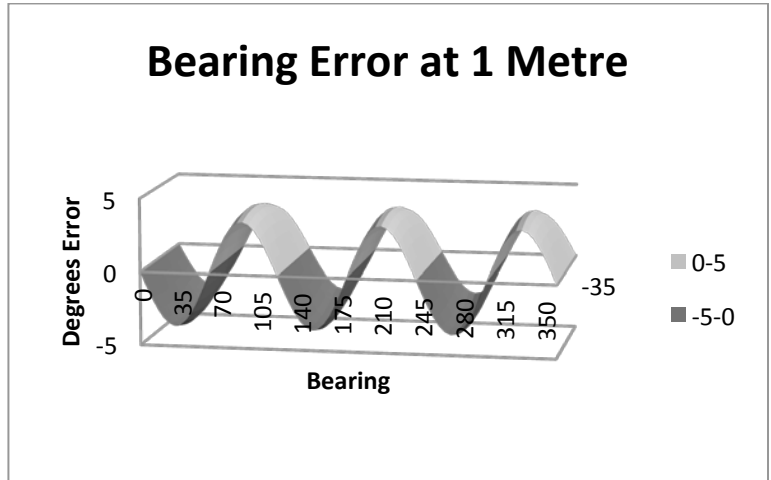

Figure 11  Bearing Error at 1 Metre

The bearing error was noted to have a value that approximates a sine wave of 3 times the periodicity of the target bearing as the bearing angle is increased. In this instance the error was found to be $\approx -4\sin(3\theta)$, where $\theta$ is the expected bearing, and was relatively free of any component introduced by elevation changes.

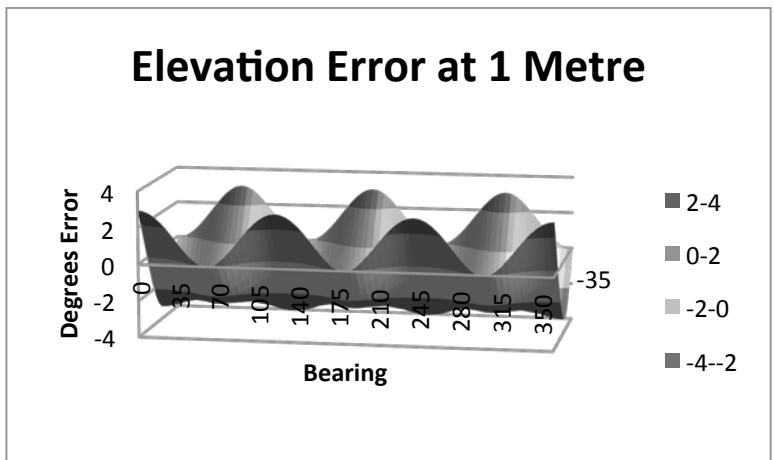

Figure 12  Elevation Error at 1 Metre

The elevation error was found to be a little more complex with a $1.5 + 1.5\cos(3\theta)$ component at -35 degrees of elevation, changing to $1.5 - 1.5\cos(3\theta)$ at 35 degrees. In this instance there is a noticeable effect caused by changes in bearing.

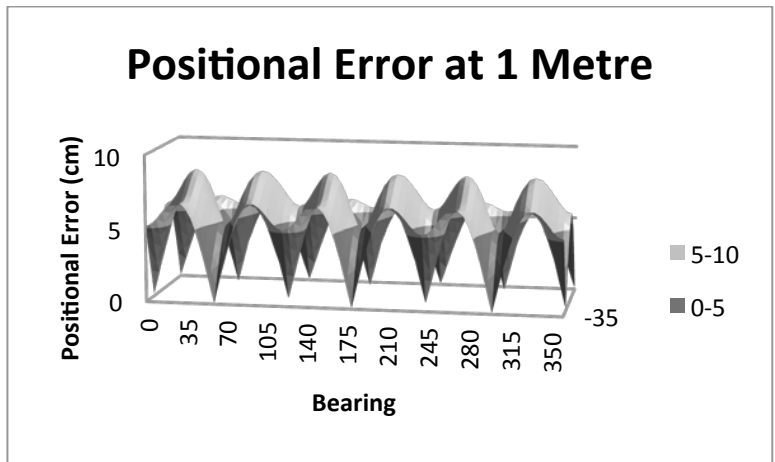

Figure 13  Positional Error at 1 Metre

The bearing and elevation measurements for the 1 metre target distance where correlated and the absolute value of the positional errors calculated. These average out at just less than 5.4 cm with a maximum error of 8.4 cm.

### 8.2. Results at 10 Metres.

A simulation was next set up for an effective target distance of 10 metres from the centroid of the tetrahedron. Simulated measurements were then made and results saved as a series of excel files from which graphs were plotted for both the bearing and elevation error components.

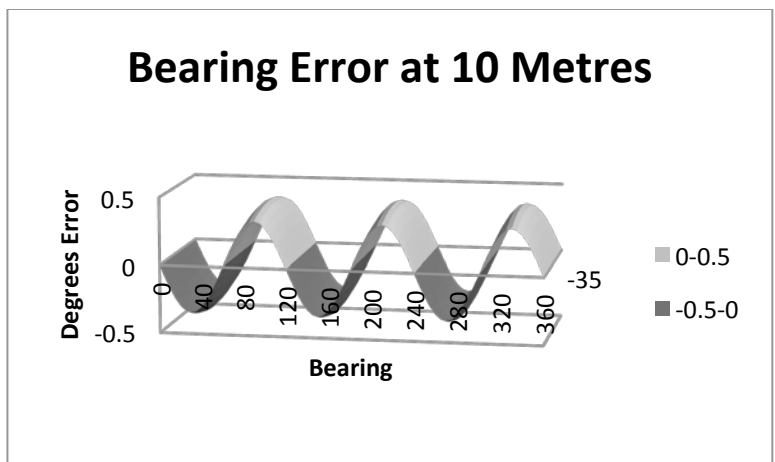

Figure 14  Bearing Error at 10 Metres

The bearing error was once more noted to have a value that approximates a sine wave of 3 times the periodicity of the target bearing as the bearing angle is increased. In this instance the error was found to be $\approx -0.4\sin(3\theta)$ and was relatively free of any component introduced by elevation changes.

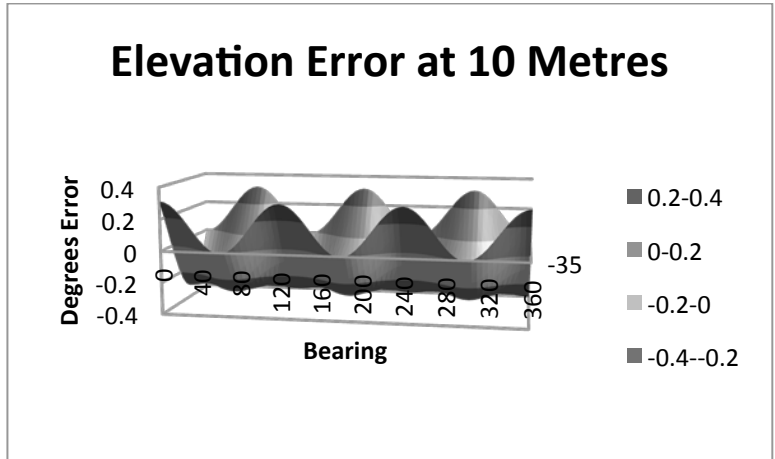

Figure 15  Elevation Error at 10 Metres

The elevation error follows a similar pattern to the 1 metre results, this time with a $0.15 + 0.15\cos(3\theta)$ component at -35 degrees of elevation, changing to $0.15 - 0.15\cos(3\theta)$ at 35 degrees. Once again there is a noticeable effect caused by changes in bearing.

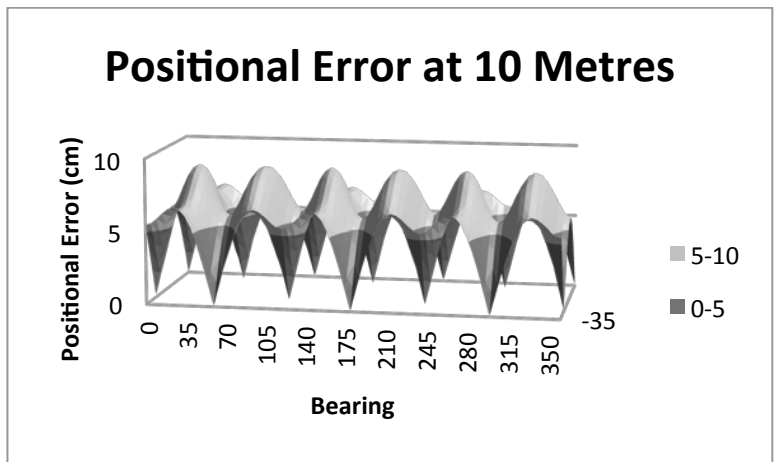

Figure 16  Positional Error at 10 Metres

The bearing and elevation measurements for the 10 metre target distance where correlated and the absolute value of the positional errors calculated. These average out at 5.61 cm with a maximum error of 8.83 cm.

### 8.3. Results at 100 Metres.

Finally a simulation was set up for an effective target distance of 100 metres from the centroid of the tetrahedron. Simulated measurements were then made and results saved as a series of excel files from which graphs were plotted for both the bearing and elevation error components.

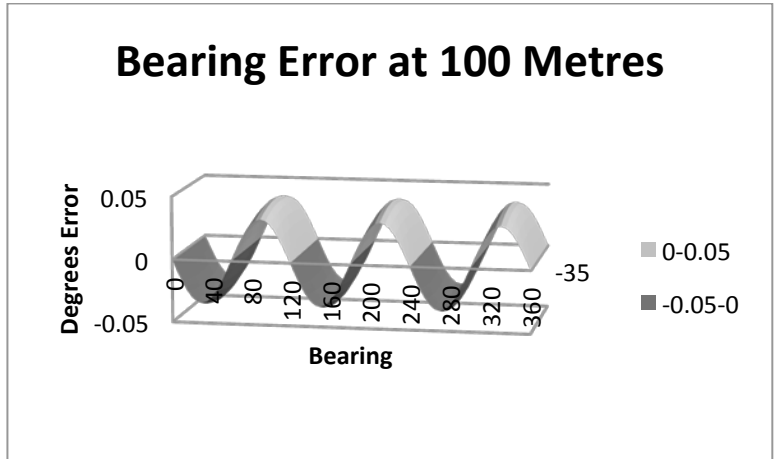

**Figure 17  Bearing Error at 100 Metres**

The bearing error was once more noted to have a value that approximates a sine wave of 3 times the periodicity of the target bearing as the bearing angle is increased.  In this instance the error was found to be $\approx -0.04\sin(3\theta)$ and was relatively free of any component introduced by elevation changes.

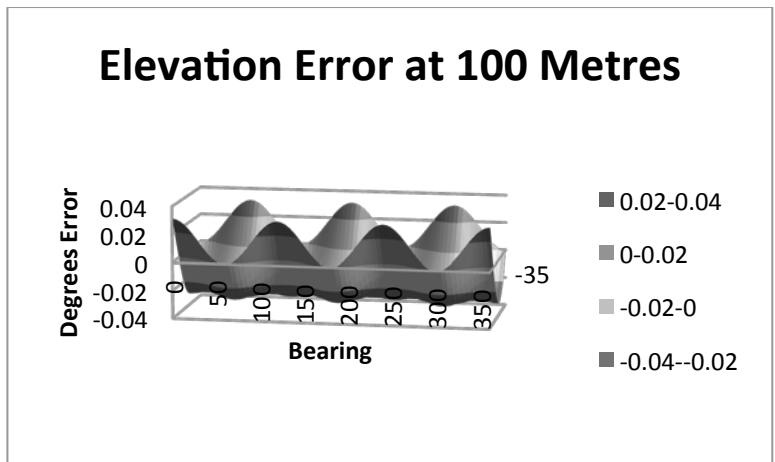

**Figure 18  Elevation Error at 100 Metres**

The elevation error, at 100 metres, follows a similar pattern to both the 1 and 10 metre results, this time with a $0.015 + 0.015\cos(3\theta)$ component at -35 degrees of elevation, changing to $0.015 - 0.015\cos(3\theta)$ at 35 degrees.  Once again there is an effect caused by changes in bearing, but this can now be seen to be decreasing as a linear function of the distance to the target.

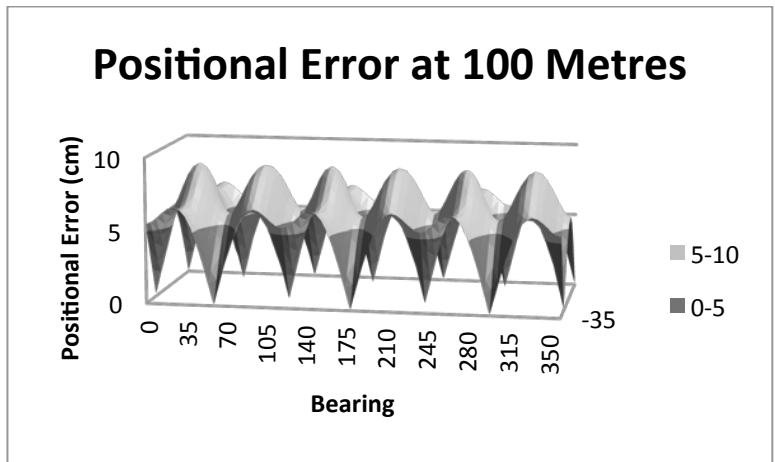
Figure 19 Positional Error at 100 Metres

The bearing and elevation measurements for the 100 metre target distance where correlated and the absolute value of the positional errors calculated. These average out at 5.61 cm with a maximum error of 8.84 cm. It can now be seen that the positional error is effectively constant regardless of distance to the target.

### 8.4. Simulation Summary

The final step was to simulate the positional errors at distances of from 1 to 1,000,000 metres at powers of 10 and the minimum, average and maximum error positional error components calculated. These are tabulated below.

Table 2 Error Components Compared to Distance

| Distance (meters) | Minimum Error (cm) | Average Error (cm) | Maximum Error (cm) |
|---|---|---|---|
| 1 | 0.07 | 5.37 | 8.41 |
| 10 | 0.08 | 5.61 | 8.83 |
| 100 | 0.08 | 5.61 | 8.84 |
| 1,000 | 0.08 | 5.61 | 8.84 |
| 10,000 | 0.08 | 5.61 | 8.84 |
| 100,000 | 0.08 | 5.61 | 8.84 |
| 1,000,000 | 0.08 | 5.61 | 8.84 |

It can be seen from the above that angular errors decrease as a linear function of distance to the target. This has the effect of producing a positional error that is unaffected by target distance. In the case of this simulation (with 50 cm antenna spacing) the maximum error is 8.84 cms.

In the case where distance to a target is known then the error component can be calculated and removed from the results. However, it is considered that the amount of error is negligible for any distance greater than, say, 10 to 100 metres.

## 9. Conclusion

This paper has presented a method of relative location by means of a tetrahedral array of antennas from which bearing and elevation may be derived. Rather than solve a typical multilateration sequence of four 3 variable simultaneous equations we, instead, directly calculate elevation and bearing information whilst assuming that our target is a point at infinity. Ranging information, in the event that it is required, may be derived from another, out of band, source.

Although this technique does have an error component we feel that the benefits of fast calculation far outweigh the inbuilt inaccuracy. Particularly, since the error component is known and can be allowed for, if required. However, given, for example, the application of an aerial robot swarm then any measurement rapidly becomes invalid due to relative movements of members of such a swarm. In this case, fast approximate updates can be viewed as much more relevant that slow updates, which although possibly accurate when the measurement was initiated may be invalid by the time that the calculation is complete.

# Glossary

| | |
|---|---|
| **CPU** | Central Processing Unit |
| **FM** | Frequency Modulation |
| **FPV** | First Person View |
| **GHz** | Gigahertz ($10^9$ cycles per second) |
| **IF** | Intermediate Frequency |
| **MHz** | Megahertz ($10^6$ cycles per second) |
| **TDM** | Time Division Multiplex |
| **TDOA** | Time Difference Of Arrival |
| **TML** | Traditional Maximum Likelihood |
| **TV** | Television |

[Figures]